\documentclass[superscriptaddress,preprint]{revtex4}
\usepackage[utf8]{inputenc}
\usepackage{amsmath}
\usepackage{amssymb}
\usepackage{natbib}
\usepackage{graphicx}
\usepackage{color}
\usepackage{bm}

\begin{document}
\title{First-principles prediction of lattice coherency in van der Waals heterostructures}
\author{Benoit Van Troeye}
\email{benoit.vantroeye@gmail.com}
\affiliation{Institute of Condensed Matter and Nanosciences, Universit\'{e} catholique de Louvain, Chemin des étoiles 8, B-1348 Louvain-la-Neuve, Belgium}
\affiliation{Department of Physics, Applied Physics, and Astronomy, Rensselaer Polytechnic Institute, Troy, New York 12180, United States}
\author{Aur\'{e}lien Lherbier}
\affiliation{Institute of Condensed Matter and Nanosciences, Universit\'{e} catholique de Louvain, Chemin des étoiles 8, B-1348 Louvain-la-Neuve, Belgium}

\author{Simon M.-M. Dubois}
\affiliation{Institute of Condensed Matter and Nanosciences, Universit\'{e} catholique de Louvain, Chemin des étoiles 8, B-1348 Louvain-la-Neuve, Belgium}

\author{Jean-Christophe Charlier}
\affiliation{Institute of Condensed Matter and Nanosciences, Universit\'{e} catholique de Louvain, Chemin des étoiles 8, B-1348 Louvain-la-Neuve, Belgium}

\author{Xavier Gonze}
\affiliation{Institute of Condensed Matter and Nanosciences, Universit\'{e} catholique de Louvain, Chemin des étoiles 8, B-1348 Louvain-la-Neuve, Belgium}
\affiliation{Skolkovo Institute of Science and Technology, Moscow, Russia}

\begin{abstract}


 The emergence of superconductivity in slightly-misaligned graphene bilayer~\cite{Cao2018} and moiré excitons in MoSe$_2$-WSe$_2$ van der Waals (vdW) heterostructures~\cite{Tran2019} is intimately related to the formation of a 2D superlattice in those systems. At variance, perfect primitive lattice matching of the constituent layers has also been reported in some vdW-heterostructures~\cite{Gong2014,Gong2015,Davies2017}, highlighting the richness of interfaces in the 2D world. In this work, the determination of the nature of such interface, from first principles, is demonstrated. To do so, an extension of the Frenkel-Kontorova (FK) model~\cite{Frank1949} is presented, linked to first principles calculations, and used to predict lattice coherency for a set of 56 vdW-heterostructures.  Computational predictions agree with experiments, when available. New superlattices as well as perfectly-matching interfaces are predicted. 

\end{abstract}
\maketitle

Combining or tuning the properties of 2D materials is now a dream come true with the advent of the so-called vdW heterostructures~\cite{Dean2010,Geim2013,Novoselov2016}, obtained by stacking different
2D materials. 
Over the last couple of years, a broad range of physical phenomena has been discovered in these systems~\cite{Novoselov2016,Zhang2016,Latini2017,Sun2015,Ross2017}, especially when the lattice mismatch and misalignment angle between the layers are small~\cite{Woods2014,Davies2017,Wilson2017,Liu2018,Zhang2017,Summerfield2018,Gong2014,Cao2018,Pan2018,Tran2019}. 
In such cases, the formation of a mesoscopic commensurate superstructure is generally
reported~\cite{Woods2014,Wilson2017,Liu2018,Zhang2017,Summerfield2018,Cao2018,Pan2018,Tran2019}, giving rise to a so-called moiré pattern of interference and to peculiar properties including unconventional superconductivity~\cite{Cao2018,Yankowitz2019}, Hofstadter butterfly~\cite{Ponomarenko2013,Dean2013,Hunt2013} or moiré excitons~\cite{Tran2019}.
The size of the underlying moiré pattern can be tuned -and consequently the electronic properties of the considered system-, not only with
the angle between the constituting layers~\cite{Woods2014}, but also with the number of layers~\cite{Wang2019}.
Experimental evidences of perfectly-matching interfaces, where at least one of the involved layer strains globally in order to match the lattice of the other layer have also been reported~\cite{Davies2017,Gong2014,Gong2015}.
These observations are reminiscent of what is achievable for epitaxial growth of III-V semiconductors~\cite{Schubert2006}, where coherent (perfectly-matching structure), 
semi-coherent (moiré pattern) and incoherent (incommensurate) interfaces can be observed depending on the lattice mismatch between the layers~\cite{Porter2009}. 

Computational material science, relying e.g.\ on Density Functional Theory (DFT)~\cite{Martin2004}, should in principle allow one to predict the nature of heterostructures.
However, determining the type of interfaces that shall be formed given two 2D materials requires to compare energetically coherent, semi-coherent and incoherent interfaces. 
While the former are easily investigated~\cite{Forster2013,Terrones2014,VanTroeye2018} with few-atom primitive cells,
semi-coherent interfaces are more challenging, as realistic repeated blocks ought to consist of hundreds or thousands of atoms. 
Eventually, the absence of a global periodicity makes incoherent interfaces inscrutable by current DFT techniques.
Recent theoretical developments try to overcome these limitations using continuum models or the  configuration space technique~\cite{Lebedev2017, Carr2018}.

For decades, the FK model has been the paradigm to study the coherency of interfaces~\cite{Frenkel1938,Frank1949,Bak1976,Woods2014}. 
This model actually yields an energy criterion for coherency determination, but such a criterion has never been applied for coherency predictions based on modern ab initio results, to the best of our knowledge. 
In this work, an extension of the FK model is derived in order to identify the global ground state of realistic 2D vdW-heterostructures from first principles. 
Along the way, additional physical features compared to the FK model are identified. Noticeably, the existence of a mean-field effect is demonstrated both theoretically and numerically. 
The predictive quality of the present model is illustrated on bilayers and periodically-repeated out-of-plane interfaces of various conventional 2D materials i.e. graphene, phosphorene, $h$-BN and Transition Metal Dichalcogenides (TMDs). Our findings agree well with the available experiments. The existence of new moiré patterns
is also predicted, like in phosphorene-$h$-BN and MoSe$_2$-WS$_2$ vdW-heterostructures. 



\begin{figure*}[htp]
 \includegraphics[width=0.95\textwidth]{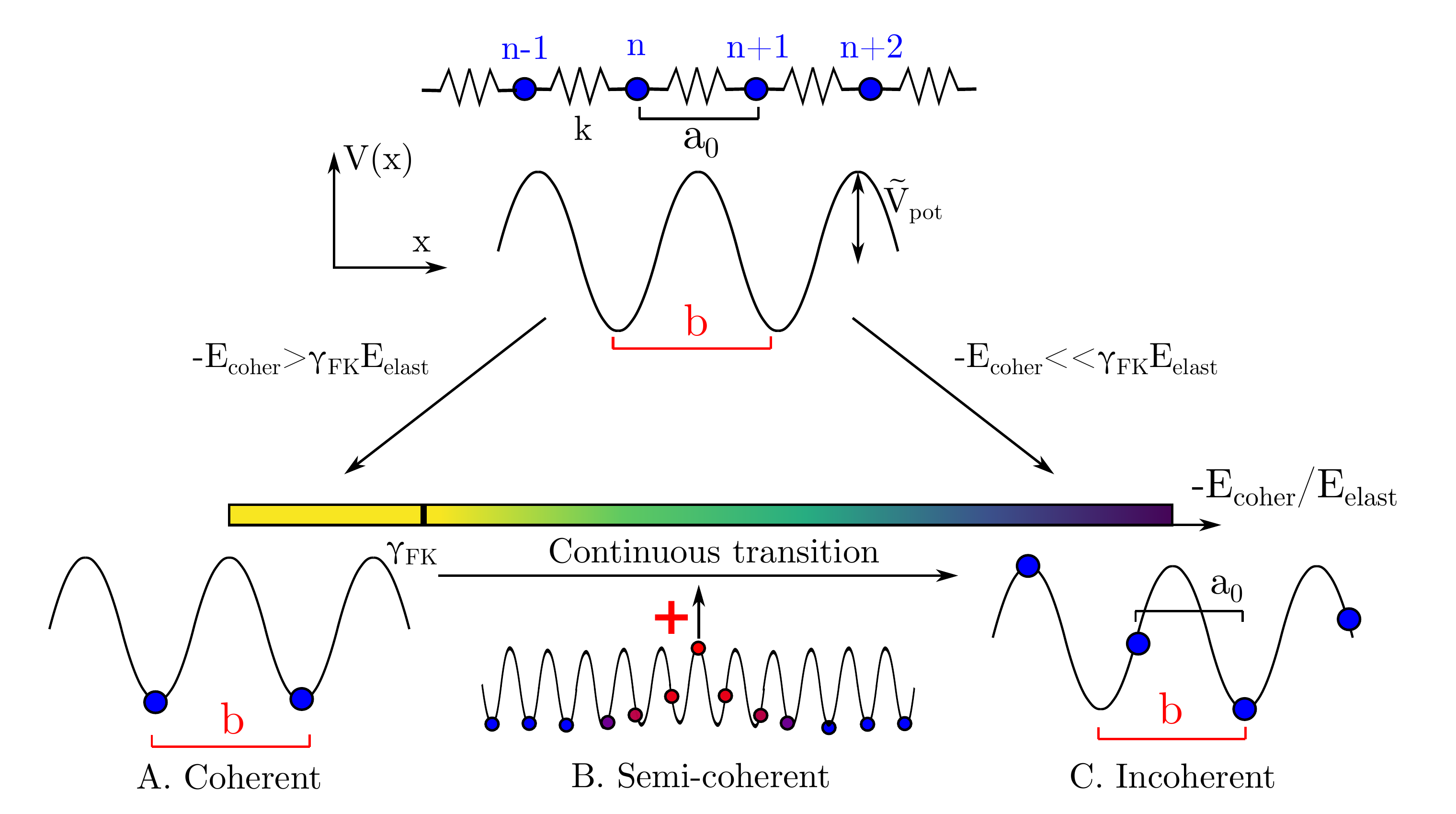}
 \caption{\label{Fig1} Schematic representation of the FK model. In a 1D atomic chain, each atom binds to its first neighbors by a spring constant $k$, and is placed in a cosine potential of amplitude $\tilde{V}_{pot}$. In order to accommodate to the potential periodicity $b$, the chain would like to strain from its initial periodicity $a_0$ but is constrained by its stiffness. If $\tilde{V}_{pot}$ is sufficiently large, then the formation of a coherent interface is favored where all the atoms fall down in the minimum valleys of the potential (case A). On contrary if the chain is rigid (case C), the chain energy is simply modulated by the average of the potential. Between these two extremes, the transition is continuous, with the progressive addition of dislocations in the system (case B). }
\end{figure*}

The classical FK model is extensively described in the S.M.-Sec. I and in 
Refs.~\onlinecite{Frenkel1938,Frank1949,Bak1982,Braun1998,Theodorou1978}. 
Briefly, it consists in a 1D periodic chain of atoms, characterized by an equilibrium lattice parameter $a_0$,
that is placed in a cosine potential:
\begin{equation}
\label{eq:Vpot}
V(x_n) = - \tilde{V}_{pot} \cos(2\pi x_n/b),
\end{equation} where $x_n$ is the position of atom $n$, $\tilde{V}_{pot}>0$ the potential amplitude, and $b\neq a_0$ its period (see Fig.~\ref{Fig1}). 
In order to accommodate to the potential, the chain tends to deform but is at the same
time constrained by its own stiffness (first-neighbor spring constant $k$). 
Within the continuum approximation~\cite{Frank1949}, the type of interface that is formed by the system is entirely determined by two quantities: first, the cost in elastic energy required
to match exactly the periodicity of the potential $E_{elast} = k(b-a_0)^2/2$, and second, the gain of potential energy resulting from the accommodation $E_{coher}=-\tilde{V}_{pot}$.
If the ratio between these two quantities overcomes in magnitude a specific threshold ($-E_{coher}\geq\gamma_{\text{FK}}E_{elast}$, with $\gamma_{\text{FK}}=\pi^2/8$~\cite{Frank1949,Theodorou1978}), the formation of a coherent interface is favored (see Fig.~\ref{Fig1}),
where each atom of the 1D chain is dragged in the bottom of a potential well. On the contrary, if the chain is rigid 
($- E_{coher}<<\gamma_{\text{FK}} E_{elast}$), it does not accommodate to the potential, and the total energy of the system is simply modified by the mean value of the potential $\langle V \rangle$ (incoherent interface), fixed to zero in Eq.~(\ref{eq:Vpot}). 
Between these two extremes,
the transition is continuous, with the progressive addition of dislocations into the system~\cite{Frank1949} (see Figs.~\ref{Fig1}
and~S1).

\begin{figure*}[htp!]
 \includegraphics[width=0.95\textwidth]{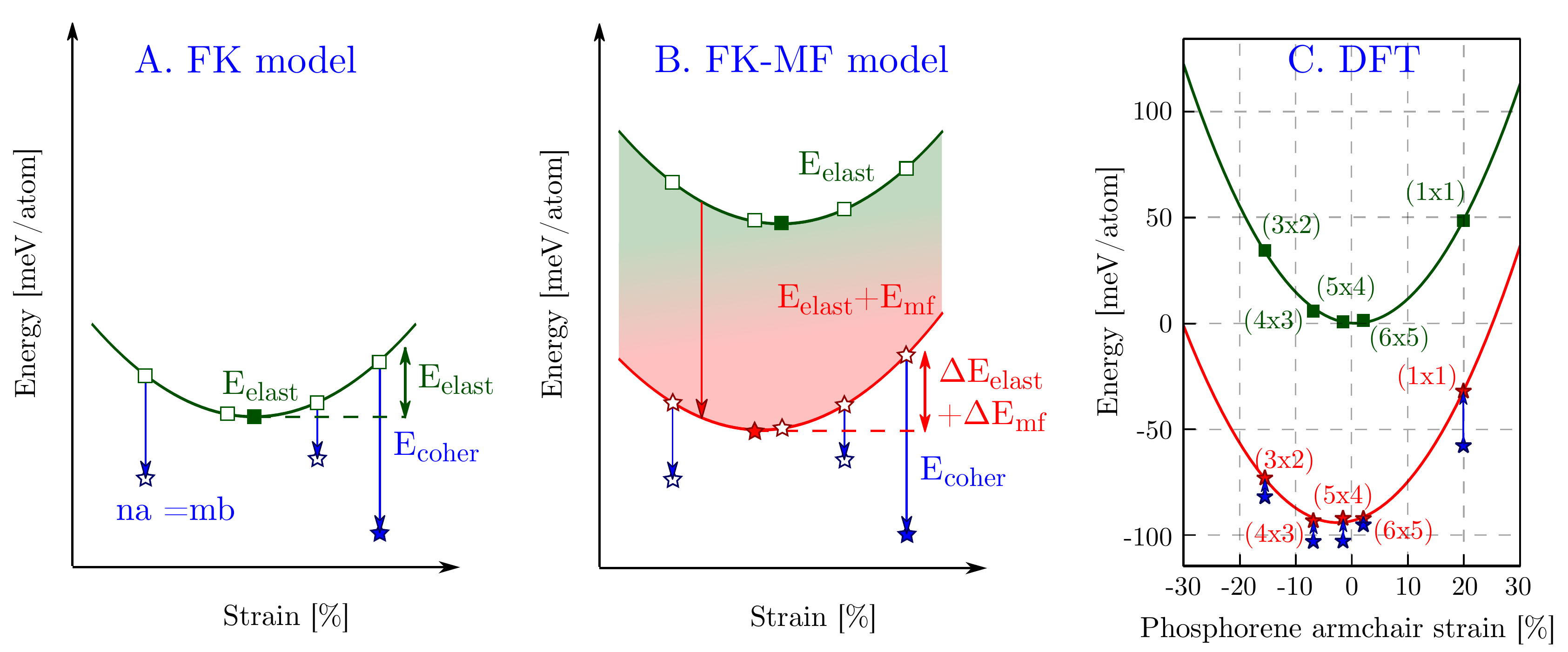}
 \caption{\label{Fig2}Schematic representation of the FK model and the extended version (FK-MF) compared to DFT calculations. The ground-state configurations of each considered case are indicated by full markers (squares or stars). \textbf{A.} Variation of the energy as function of the (homogeneous) strain of the 1D chain in the FK model. In the isolated case, the energy varies quadratically (green squares). When the lattice of the 1D chain becomes commensurate with respect to the substrate (na = mb), a discrete gain of energy is achieved, which differs for each commensurate structure (blue arrow). This gain may or may not overcome the elastic cost to reach that strain state. \textbf{B.} In addition to the (continuous) elastic and (discrete) coherency energies already present in the FK model, a continuous mean-field term, corresponding to the averaged interaction between the layers, has to be added in order to properly describe the physics of the considered interface problem (red stars). This mean-field effect leads to a shift of the equilibrium lattice of the constituent layers with respect to their isolated forms (the strain of the green full square and red full star are not identical). \textbf{C.} DFT computations illustrating the correspondence with {\it ab initio} calculations in the phosphorene-MoSe$_2$ vdW-heterostructure case. The mean-field interaction is extracted using Eq.~(\ref{eq:meanfield}). The mean-field interaction leads to a shift of the equilibrium lattice parameter of phosphorene compared to its free-standing form.  }
\end{figure*}

In principle, the type of interface is entirely determined by the $E_{coher}$ and $E_{elast}$ fundamental quantities. Still, higher-order coherent interfaces
should also be considered~\cite{Aubry1979}, corresponding to the cases where $na =mb$ ($n,m$ integers) (see S.M. and Fig.~S2). Hence, the ground state of the problem has to be searched in this whole set of possible configurations,
as illustrated schematically in Fig.~\ref{Fig2}A.


As it stands, the FK model allows one to simulate the formation of (semi-)coherent interfaces between an epitaxial layer (the 1D chain) and a rigid substrate (the potential). The objective is now to extend this classical picture to 
the realistic case of an interface composed of two different 2D materials, both characterized by their own set of properties (lattice parameters, elastic constants, ...).
A similar coherency criterion can indeed be formulated for such realistic systems (see S.M.-Sec.~II), although it requires the introduction of some additional features, as described now. We denote this extended model as the Frenkel-Kontorova-mean-field (FK-MF) model.

First, each 2D material creates a potential that is felt by the other layers, which can in turn deform in order to accommodate to the incurred potential (see Fig.~S5). 
Similarly to the FK model, lattice matching yields a gain in interlayer potential energy,
at the cost of elastic energy; still now both materials can strain in order to form a coherent interface. For sake of simplicity, the layers are supposed to be aligned along the same crystalline 
directions such that a common conventional cell can be constructed. 
To do so, the conventional orthogonal cells of the first and second layers (referred hereafter as the building cells) are repeated, respectively $n^{\text{(1)}}_{\alpha}$ and $n^{\text{(2)}}_{\alpha}$ times, along the $\alpha$ in-plane direction. Then, the two layers must be strained to match each other, with respective strains $\epsilon^{\text{(1)}}_{\alpha}$ and
$\epsilon^{\text{(2)}}_{\alpha}$ in Voigt notation. Note that, since the layers are aligned, there is no shear strain incurred by the layers. In the linear elastic regime, this leads
 to an increase of elastic energy per atom (see S.M.-Secs.~II to~IV):

\begin{equation}
\begin{split}
 E_{elast} = \frac{1}{N_{at}} \left[\frac{A^{\text{(1)}}_0}{2} \prod_{\gamma} n^{\text{(1)}}_{\gamma}  \sum_{\alpha,\beta} \tilde{c}^\text{(1)}_{\alpha\beta} \epsilon^{\text{(1)}}_{\alpha} \epsilon^{\text{(1)}}_{\beta} \right. \\ 
 \left. + \frac{A^{\text{(2)}}_0}{2} \prod_{\gamma} n^{\text{(2)}}_{\gamma} \sum_{\alpha,\beta} \tilde{c}^\text{(2)}_{\alpha\beta} \epsilon^{\text{(2)}}_{\alpha} \epsilon^{\text{(2)}}_{\beta} \right], \label{eq:dEtotal} \\ 
\end{split}
\end{equation}
where $N_{at}$ is the number of atoms in the (higher-order, if $n_{\gamma}^{\text{(1,2)}}>1$) coherent interface, $A^{\text{(1)}}_0$ and $A^{\text{(2)}}_0$ are the areas of the undeformed conventional building cells of the first
and second layers, respectively,
$\tilde{c}^\text{(1)}_{\alpha\beta}$ and $\tilde{c}^\text{(2)}_{\alpha\beta}$ are their elastic constants (per surface area) expressed in Voigt notation, $\alpha=1,2$, $\beta=1,2$ and
$\gamma=1,2$.
The way the strains are split between the layers is determined by, first enforcing lattice matching (see Eq.~(S75)) and by cancelling the total tension in the layers (see Eq.~(S77)).

Second, as soon as a coherent interface is created between two 2D materials, there exists an optimal-stacking configuration that minimizes the interlayer interaction. 
By comparison with an incoherent interface, where the interlayer interaction is averaged over the whole translation landscape, there is a gain in energy to form a (semi-)coherent interface, counterbalanced by the elastic energy cost to form it. This (discrete) gain of energy
is referred here as the coherency energy $E_{coher}$, and simply corresponds to
\begin{equation}
 E_{coher} = \frac{1}{N_{at}} \left[ E^{min}_{tot} - \left< E_{tot} \right> \right], \label{eq:ECoher1}
\end{equation}
where $E^{min}_{tot}$ is the total energy of the supercell spanning the coherent interface in the optimal stacking configuration, and $\left< E_{tot} \right>$ is
the total energy of the same supercell averaged over the translational landscape i.e.
\begin{equation}
 \left< E_{tot} \right> = \frac{1}{A_{0,S}} \int_{A_{0,S}} E_{tot}(\bm{\tau}) \, \mathrm{d}\bm{\tau}, \label{eq:ECoher2}
\end{equation}
where $\bm{\tau}$ is an in-plane vector corresponding to a rigid translation of the first layer with respect to the second, and $A_{0,S}$ is the area formed by the supercell. 

Third, a rigorous treatment of the problem, as provided in Eqs.~(S34) to~(S50), reveals 
the importance of the averaged interlayer interaction, in contrast to the FK model where its contribution vanishes ($\langle V \rangle=0$). This term, referred in the following as the mean-field interaction (energy contribution $E_{mf}$), corresponds to the stacking energy~\cite{VanTroeye2018} averaged over the whole interface or over all in-plane translations in the case of a coherent interface (see Eq.~(S47)).
This mean-field term does not only modify the way the strains are split between the layers in the case of a coherent interface,
but also affect the ground-state lattice parameters and the elastic constants of the involved layers in the incoherent case compared to their free-standing forms (see Figs.~S4 and~\ref{Fig2}B).

The threshold separating the coherent and semi-coherent regimes has thus to be adapted compared to the plain FK model to take into account the mean-field interaction and its consequences. 
As discussed
in the S.M.-Secs.~II and~III, the formation of a coherent interface becomes favorable when
\begin{equation}
 -E_{coher}\geq\gamma \left[ \Delta E_{elast}+ \Delta E_{mf} \right], \label{criterion}
\end{equation}
where $\gamma$ is a constant, whose value depends on the shape of the interlayer potential ($\gamma = \gamma_{\text{FK}} = \pi^2/8 \approx 1.23$ for single sine or cosine component).
The differences in elastic energy and mean-field energy $\Delta E_{elast}+ \Delta E_{mf}$ are to be evaluated with respect to those at the lattice parameters that minimize the energy of the incoherent interface, instead of those of the isolated subsystems, see Fig.~\ref{Fig2}B. 


From a practical point-of-view, the ingredients required to predict the type of interface are the elastic, coherency and mean-field energies. Both elastic and coherency energies can easily be computed
based on the relaxed coherent interface, obtained for example in DFT, even for higher-order coherent interfaces (see S.M.-Sec.~IV). One still has to define a strategy in order to estimate the mean-field
interaction, which requires in principle to compute the variation of the interlayer energy with strain averaged over the whole interface.

An analytical expression for the mean-field interaction could be derived using perturbation theory (see S.M.-Subec.~II.D), but this contribution can equally well be estimated numerically based on several DFT computations,
as follows.
For coherent interfaces, the mean-field energy is available as
\begin{equation}
\label{eq:meanfield}
E_{mf} =
E_{tot}^{\text{(1+2)}}
-E_{tot}^{\text{(1)}}
-E_{tot}^{\text{(2)}}
- E_{coher}, 
\end{equation}
where
$E_{tot}^{\text{(1+2)}}$ is the total energy of the
interface supercell, while $E_{tot}^{\text{(1)}}$
and $E_{tot}^{\text{(2)}}$ are the total energies of the
separate subsystems in the same strain state,
see Eqs.~(S48) and (S80).
In order to extract the variation of the mean-field energy with strain, different coherent interfaces have to be considered and relaxed in DFT. 
Then, the mean-field energy, based on Eq.~(\ref{eq:meanfield}), is simply interpolated with a low-order polynomial.

To illustrate this approach, DFT computations have been performed on different coherent interfaces, i.e. perfectly-matching or higher-order ones, constituted of MoSe$_2$ and phosphorene in the aligned case (zigzag
on zigzag) and periodically-repeated out-of-plane to increase the magnitude of the interlayer interactions compared to a simple bilayer (see S.M.-Sec.~V).
These two materials have a good lattice matching along the zigzag direction ($<$0.3\%), while the elasticity window of phosphorene in
the armchair direction is sufficiently large (up to $\sim$ 20$\%$)~\cite{Peng2014} for the small-size structures with large strain states investigated in this work to be still meaningful. 
The ``pure'' elastic contribution was computed based on Eq.~(\ref{eq:dEtotal}) using the equilibrium lattices and elastic
constants of the isolated layers, as well as the relaxed lattices of the coherent structures, as predicted by DFT (see the Methodology and S.M.-Sec.~V). The coherency energy was calculated fixing all the other degrees of freedom. Eventually, the mean-field energy was evaluated following Eq.~(\ref{eq:meanfield}).

The different results are shown in Fig.~\ref{Fig2}C as function of the strain incurred by phosphorene along the armchair direction (see Fig.~(S7) for the exact path in the strains hyperspace). As one can see, the minimum of the elastic plus mean-field term (red curve) is shifted with respect to the pure elastic case (green curve), in agreement with our theoretical developments. This is further confirmed by looking
at the averaged strain incurred by the layers as shown in Fig.~S7. 
Note that the $E_{elast}+E_{mf}$ term is as expected dominated by $E_{elast}$ since the mean field interaction between layers is weak. 
The most noticeable changes happen close to the energy minimum, where the linear terms dominates over the quadratic ones; in first approximation, the mean-field mostly modifies
the lattice parameters of the involved layers. A similar analysis has been performed along the MoSe$_2$ armchair direction (see Fig.~S10) and reveals negligible variation of the lattice parameter
with respect to its isolated counterpart. MoSe$_2$ is indeed stiffer than phosphorene along the armchair direction. In general, the most noticeable effects are thus expected to occur in phosphorene, the softest (along the armchair direction) 2D material effectively 
exfoliated~\cite{Choudhary2018} up to date, and to be imperceptible in the case of stiffer materials like graphene.
This mean-field effect also explains why the armchair lattice parameters of phosphorene are predicted to differ from its bulk counterpart~\cite{Shulenburger2015,Wei2018,VanTroeye2018}. Further discussions on the phosphorene-MoSe$_2$ interface can be found in S.M.-Sec.~V.


Once the physics behind interface formation has been clarified, the FK-MF model can be used to predict the type of interface that shall be formed for a given pair of 2D materials.
Eight 2D materials are considered herewith: graphene, $h$-BN, phosphorene, MoS$_2$, MoSe$_2$, MoTe$_2$, WS$_2$ and WSe$_2$ (see Tab.~S2 for a list of their respective properties). 
For the coherency investigation, only perfectly-matching interface between 
two materials are selected. Hence, even if an incoherent interface is predicted here, the formation of a moiré pattern corresponding to higher-order coherent interfaces is still possible experimentally. The aligned -or anti-aligned case for TMDs heterostructures-
cases were considered. When the crystal systems of the chosen 2D materials do not match, the coherency has been investigated in a specific lattice direction, described in the S.M.-Sec.~V. Both the
bilayer and periodically-repeated out-of-plane vdW-heterostructures cases were considered. Further information can be found in S.M.-Sec.~VI.

As discussed previously, the elastic and coherency energies are relatively straightforward to compute, in contrast to the mean-field energy. 
To overcome this difficulty, we exploit the fact that this mean-field term impacts as well the bulk phase of the considered 2D materials, like graphite and black phosphorus, for which its effect can effectively be computed~\cite{VanTroeye2018}.
For example, in black phosphorus, the relative change of armchair lattice parameter is predicted to be -3.7~\% going from monolayer to the bulk using~\cite{VanTroeye2018}, to be compared to -2.5~\% derived previously in this work for phosphorene-MoSe$_2$.
A reasonable approximation for the mean-field interaction is thus to consider that the ground-state lattice parameter of the involved layers in the incoherent state are no more the ones of their isolated forms for the estimation of the elastic energy, but instead the ones of their bulk counterparts 
(e.g. graphite for graphene) for periodically-repeated vdW-heterostructures, or of their bilayer counterparts (e.g. graphene bilayer for graphene) for bilayer vdW-heterostructures. 
This allows one to use another form of the coherency criterion, Eq.~(S52).
As introduced before, significant effects are only observed for phosphorene while the change of lattice parameter is negligible in all the other considered materials ($<$0.2\% change). The interface nature is not only determined by the elastic, mean-field and coherency energies, but also by the $\gamma$ constant which depends on the shape of the interlayer potential. 
In consequence, the translational landscapes of some reference vdW-heterostructures are shown in Fig.~S12. 
Although they are generally complex, these translational landscapes can in good approximation be described along the high-symmetry lattices by a single sine (or cosine). 
This incites us to approximate $\gamma=\gamma_{\text{FK}} = \pi^2/8 \approx 1.23$ in Eq.~(\ref{criterion}) to assess the interface nature in what follows. 

\begin{figure*}[htp!]
 \includegraphics[width=1\textwidth]{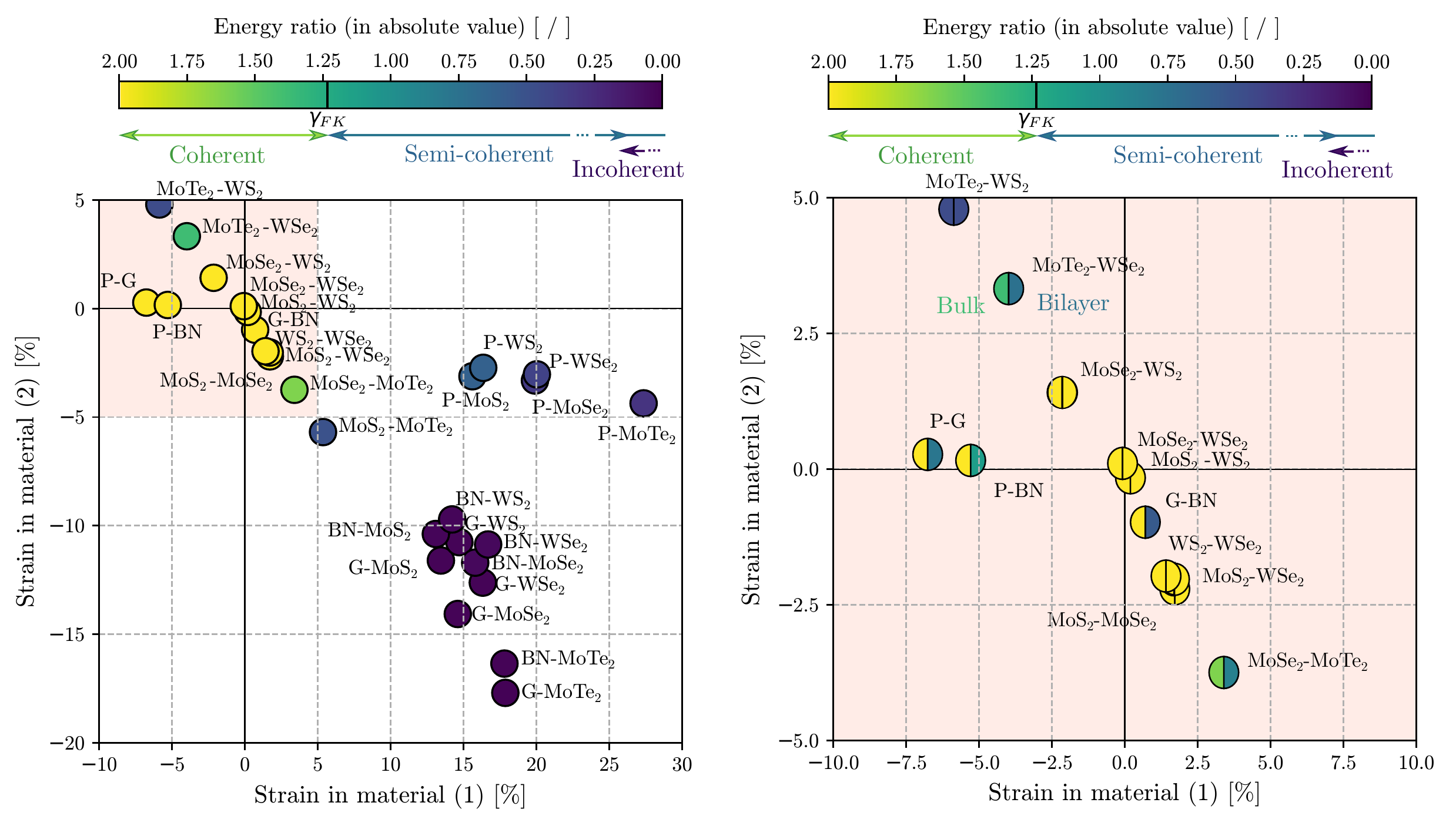}
\caption{\label{Fig3}Prediction of lattice coherency in vdW-heterostructures: computed energy ratio for 28 pairs of 2D materials. A. For periodically-repeated out-of-plane vdW-heterostructures. B. corresponds to a zoom on the interesting structures of A with the highest energy ratio: for them, both monolayer stackings, forming the periodically-repeated vdW-heterostructures (left part of a disk) and bilayers (right part of a disk) are investigated. The energy ratio of the bilayer is always larger than the vdW-heterostructure one.}
\end{figure*}

The interface predictions are shown in Fig.~\ref{Fig3}A for periodically-repeated vdW-heterostructures, and for bilayers in Fig.~\ref{Fig3}B. The numerical values are also presented in Tab.~S4. 
Note that, when their interfaces were predicted as semi-coherent or incoherent in the periodically-repeated case, the coherency and elastic energies of their bilayer counterpart were not computed. Indeed, their corresponding elastic energy remains approximately the same compared to the periodically-repeated case, while the coherency energy decreases approximately by a factor 2. Those interfaces are therefore always still semi-coherent or incoherent.
The most interesting systems are contained within a $\sim10\%$ window centered at the origin. Quite a few bilayer systems are predicted to be semi-coherent (graphene-hBN, 
hBN-phosphorene, graphene-phosphorene, MoTe$_2$-WSe$_2$, MoSe$_2$-MoTe$_2$), while most bilayer interfaces containing TMDs are predicted to be coherent.
Interestingly, most of the semi-coherent bilayer systems turn coherent moving to periodically-repeated out-of-plane heterostructures.

Some of the considered systems have actually been synthesised either using the pick-and-lift
technique~\cite{Dean2010,Davies2017,Wilson2017,Liu2018} 
or by epitaxial growth~\cite{Davies2017,Gong2015,Zhang2017,Gong2014}. 
In some cases, both semi-coherent and coherent interfaces have been reported depending on the synthesis method~\cite{Wilson2017,Gong2015,Woods2014,Davies2017}. We expect the epitaxially grown interfaces to be closer to the theoretical ones, since such a method allows the epitaxial layer to
accommodate to the substrate without being limited by energy barriers (i.e. the energy required to extract the dislocations from the system~\cite{Frank1949}). 
In consequence, our results will be compared in the following
to the experimental grown interface if available. Even so, a subtle difference between our approach and experiments remains, regarding the influence of the additional layers
constituting the substrate. They are not considered in the model while they somehow constrain the way the layers deform to form the (semi-)coherent interface. 

Taking these considerations into account, our predictions compare well with experiments: MoS$_2$-WS$_2$ and MoSe$_2$-WSe$_2$ bilayers are correctly predicted as coherent~\cite{Gong2014,Gong2015}, similarly to
graphene-hBN and phosphorene-graphene bilayers (semi-coherent)~\cite{Woods2014,Liu2018}. To the best of our knowledge, no moiré pattern was ever reported for graphene-MoS$_2$ 
vdW-heterostructure~\cite{Pierucci2016}, in agreement with our calculations: the gain in coherency  energy is too small to counterbalance the enormous elastic cost required to match the lattices of 
graphene and MoS$_2$. Finally, although they indicate correctly that lattice accommodation should occur in MoS$_2$-WSe$_2$ interfaces, our computations incorrectly predict this system to be coherent,
while a semi-coherent interface is reported experimentally~\cite{Pan2018,Zhang2017}. We believe that this discrepancy originates from the flat wells of its interlayer potential, see Fig.~S12, which leads to a $\gamma$ that differ from $\gamma_{FK}$ in a non-negligible manner.


To wrap up, the FK-MF model linked to first-principles calculations is a powerful new strategy to investigate vdW-heterostructures from an {\it ab initio} perspective. It relies on a few DFT quantities 
-lattice parameters and elastic constants of the separate systems, energy of relaxed coherent interface and translational landscape between the layers, etc.-,
combined with a simple energy criterion. 
This yields a quick determination of the interface that shall be formed during experimental growth of vdW-heterostructure by comparing simple functions of the above quantities. 
The FK-MF model has been being applied to various 2D materials. 
The results compare well with available experimental results, while
lattice coherency is predicted in several new systems not yet investigated experimentally.

\subsection*{Acknowledgment}
A. Lherbier and S. M. Dubois contributed equally to this work. The authors acknowledge technical help from J.-M.~Beuken and M. Giantomassi. This work has been supported by the FRS-FNRS through a FRIA Grant (B.V.T.) and two research projects (N$^\circ$T.1077.15 \& T.0051.18); the Communauté française de Belgique through the BATTAB project (ARC 14/19-057) and the \textit{``3D nanoarchitecturing of 2D crystals''} project (ARC 16/21-077); the Région Wallonne through the BATWAL project (N$^\circ$1318146); the European Union's Horizon 2020 research and innovation program (GrapheneFlagship Core1 - N$^\circ$696656 and Core2 - N$^\circ$785219). Computational resources have been provided by the supercomputing facilities of the UCLouvain (CISM) and the Consortium des Equipements de Calcul Intensif en Fédération Wallonie Bruxelles (CECI) funded by the Fonds de la Recherche Scientifique de Belgique (FRS-FNRS) under convention 2.5020.11. The present research benefited from computational resources made available on the Tier-1 supercomputer of the Fédération Wallonie-Bruxelles, infrastructure funded by the Walloon Region under the grant agreement N$^\circ$1117545.

\subsection*{Method and computational details}
\textit{Ab initio} computations are based on Density Functional Theory~\cite{Martin2004} as implemented in the \textsc{Abinit} software package~\cite{Abinit2005,Abinit2009,Gonze2016}. The exchange-correlation potential comes from the GGA-PBE functional~\cite{Perdew1996} with Grimme's DFT-D3 dispersion corrections to include the long-range e$^-$-e$^-$ correlation~\cite{Grimme2010,VanTroeye2016}.
The cut-off radius for the coordination number, required for the dispersion corrections, is set to 105~\AA $\,$  and only pairs contributing for more than $10^{-12}$ Ha are taken into account. Calculations are based on a planewave basis set and ONCVPSP norm-conserving pseudopotentials \cite{Hamann2013} from the PseudoDojo project \cite{VanSetten2018} thus including multiple angular projectors. 
In the case of the MoSe$_2$-phosphorene vdW-heterostructure, a planewave cut-off energy  of 54~Ha and a 10$\times$10$\times$1 Monkhorst-Pack wavevector grid~\cite{Monkhorst1976} are found sufficient for convergences of the ground-state energy of phosphorene and MoSe$_2$ building blocks up to 1~meV/atom. The k-point mesh is then adapted as function of the superlattice dimensions. 10 k-points are used to sample the out-of-plane direction for the periodically-repeated vdW-heterostructure. The structure of the constructed superlattices are then optimized such as the maximum force incurred by the atoms is smaller than $10^{-5}$ Ha/Bohr. Elastic constants are estimated using Density Functional Perturbation Theory~\cite{Gonze1997,Gonze1997b,Hamann2005,Hamann2005b,Hamann2005c,Gonze2005a,VanTroeye2017}. For the lattice coherency prediction, the same cut-off energy was found sufficient for convergence purposes, but denser k-point meshes were required for some materials, like graphene and hBN (up to 18$\times$18$\times$1). The postprocessing of the \textsc{Abinit} outputs to feed the FK-MF model was performed using \textsc{python} and \textsc{Abipy}~\cite{Gonze2016}; the corresponding scripts and data are provided in the S.M. through Jupyter Notebooks. 

\end{document}